\documentclass[11pt, 5p,times,twocolumn]{article}
\pdfoutput=1
\usepackage[a4paper, total={7.5in, 10in}]{geometry}
\usepackage{amssymb}
\usepackage{amsmath}
\usepackage{amsfonts}
\usepackage{algorithmic}
\usepackage{graphicx,color}
\usepackage{textcomp}
\usepackage{xcolor}
\usepackage{verbatim}
\usepackage{colortbl}
\usepackage{booktabs}
\usepackage{caption}
\usepackage{subcaption}
\usepackage{xurl}
\usepackage[normalem]{ulem}
\usepackage{tabularray}
\usepackage{siunitx}
\usepackage{flushend}
\usepackage{comment}
\usepackage{adjustbox}
\usepackage{makecell}
\usepackage{multirow}
\usepackage{rotating}
\usepackage{enumerate}
\usepackage{xcolor}
\usepackage{lscape}
\usepackage{comment}
\usepackage{booktabs} 
\usepackage{adjustbox}
\usepackage{float}
\usepackage[affil-it]{authblk}
\usepackage{blindtext}
\usepackage{abstract}
\begin{document}
\title{Security-Aware Availability Modeling of a 5G-MEC System}
\author[1]{Thilina Pathirana%
\thanks{This work was supported by the Norwegian Research Council through the 5G-MODaNeI project (no. 308909).}}
\affil[1]{University of Stavanger}
\author[1]{Gianfranco Nencioni}
\author[2]{Ruxandra F. Olimid}
\affil[2]{University of Bucharest}
\date{June 2024}
\twocolumn[
\begin{@twocolumnfalse}
\maketitle
\begin{abstract}
Multi-access Edge Computing (MEC) is an essential technology for the fifth generation (5G) of mobile networks. MEC enables low-latency services by bringing computing resources close to the end-users. The integration of 5G and MEC technologies provides a favorable platform for a wide range of applications, including various mission-critical applications, such as smart grids, industrial internet, and telemedicine, which require high dependability and security. Ensuring both security and dependability is a complex and critical task, and not achieving the necessary goals can lead to severe consequences. Joint modeling can help to assess and achieve the necessary requirements. Under these motivations, we propose an extension of a two-level availability model for a 5G-MEC system. In comparison to the existing work, our extended model (i) includes the failure of the connectivity between the 5G-MEC elements and (ii) considers attacks against the 5G-MEC elements or their interconnection. We implement and run the model in M\"{o}bius. The results show that a three-element redundancy, especially of the management and core elements, is needed and still enough to reach around 4-nines availability even when connectivity and security are considered. Moreover, the evaluation shows that slow detection of attacks, slow recovery from attacks, and bad connectivity are the most significant factors that influence the overall system availability. 
\end{abstract}
\vspace{1.5cm}
\end{@twocolumnfalse}
]
\saythanks

\section{Introduction}

Multi-access Edge Computing (MEC), a pivotal technology in the landscape of the fifth generation (5G) of mobile networks, strategically situates computing resources and services at the network edge. The 5G-MEC systems bring several benefits, including reduced latency and enhanced context awareness, which are vital for real-time data-intensive applications. By offloading tasks traditionally handled in cloud data centers to the edge nodes, MEC effectively reduces the traffic volume through the network core, alleviating congestion and speeding up response times. Consequently, MEC helps to achieve the latency and reliability requirements of critical 5G applications, such as autonomous vehicles, telemedicine, augmented reality, virtual reality, and smart city infrastructures~\cite{ITU2083-0,tran2017collaborative}.

However, the shift towards edge computing also complicates the network architecture. The introduction of additional nodes and the distribution of computational tasks across the network perimeter increase the system complexity, posing significant challenges in maintaining security and dependability~\cite{roman2018mobile,shi2016edge}. In this context, dependability pertains to the reliability and availability of the system, focusing on resilience to unintentional disruptions such as hardware failures or software bugs. On the other hand, security addresses the system protection against intentional attacks aiming to, e.g., disrupt the service or steal the data. Ensuring high levels of both security and dependability is crucial. Otherwise, the functionality, performance, and trustworthiness of 5G might suffer~\cite{tran2017collaborative}. Therefore, it is essential for 5G-MEC systems not only to perform efficiently under normal conditions but also to maintain robustness and accessibility in the face of both unexpected faults and deliberate security threats.

\subsection{Motivation}

Designing highly dependable and secure systems can pose challenges and potentially escalate costs~\cite{dep:concepts}. It is crucial to understand that while security and dependability are distinct, they are interrelated. Integrating both can offer a more comprehensive system model, revealing characteristics that might be overlooked if only one domain is considered~\cite{jonsson1992integration}.

\textit{Availability}, the readiness to provide accurate service, is integral to dependability and security. In Ultra-Reliable Low-Latency Communication (URLLC), the service must be available 99.999\% of the time, achieving a five-nines availability~\cite{five9s}. System-wide availability is paramount in a 5G-MEC setting, especially with mission-critical applications, and it is dependent on both faults and attacks. 

While some existing studies offer valuable insights, work has yet to address a complete 5G-MEC system security and dependability modeling. Our study aims to bridge this gap by proposing an availability model encompassing the various security and dependability facets of a 5G-MEC system.

\subsection{Contribution}

The two-level availability model introduced in~\cite{pathirana2023availability} offers a holistic perspective on system availability by accounting for the failure of various components (e.g., hardware, software) within each element of the 5G-MEC system. However, the model overlooks the possible attacks against the system availability. To address this, we expand the model to encompass the system unavailability caused by attacks.
As a second contribution, we extend the model to consider the failure of the connectivity between elements (e.g., the connectivity between the user equipment and the radio unit), another aspect that the reference paper \cite{pathirana2023availability} entirely ignores. More precisely, in addition to \cite{pathirana2023availability}, our model looks into (i) the unavailability of the connectivity between the elements of the considered 5G-MEC system caused by \textit{unintentional events} and (ii) the unavailability of the elements themselves or the connectivity between the elements, caused by \textit{intentional adversaries}. Unlike other previous works that target specific system components, our model (same as the model in \cite{pathirana2023availability}) considers dependability and security impact on all the elements of the 5G-MEC system, including management. 

We define the proposed model in M\"{o}bius~\cite{mobiusweb}, a tool designed for modeling, analyzing, and evaluating complex systems, particularly in reliability, availability, security, and performance, and commonly used in the related literature. We further use the proposed model to evaluate the impact of security and connectivity issues on the overall availability of the system. We investigate the best redundancy configuration to maximize availability. We also identify the critical elements and components that may lead to a disruption of the 5G-MEC system.

\subsection{Outline}

The rest of the paper is structured as follows. Section~\ref{sec:related_works} briefly presents the related works. Section~\ref{sec:background} gives the necessary background, including the 5G-MEC general architecture, the considered implementation, and some notes on attacks against availability in 5G-MEC systems.
Section~\ref{sec:dep_sec_modeling} presents the basic notions to understand the proposed model, as well as the main references we base our model on. Section~\ref{sec:model} introduces the proposed availability model. Section~\ref{sec:eval} evaluates and discusses the 5G-MEC system availability using the model and considering diverse system configurations and parameter variations. Section~\ref{sec:conclude} concludes. An alphabetical list of acronyms is given in Table~\ref{tab0} for the reader's convenience.

\begin{table}[t!]
\caption{List of Acronyms}
\begin{center}
\adjustbox{width=0.45\textwidth,keepaspectratio}{
\begin{tabular}{ll}
\toprule

5G	&	Fifth Generation of mobile networks\\
5GC	&	5G Core\\
APP	&	Application\\
COTS	&	Commercial-Off-The-Shelf\\
CU	&	Central Unit\\
DU	&	Distributed Unit\\
ETSI &  European Telecommunications Standards Institute\\
FT	&	Fault Tree\\
FW	&	Firmware\\
HW	&	Hardware\\
MANO	&	Management and Orchestration\\
MEC	&	Multi-access Edge Computing \\
MEH	&	MEC Host\\
MEO	&	MEC Orchestrator\\
MEP	&	MEC Platform\\
MEPM	&	MEC Platform Manager\\
NFV	&	Network Function Virtualization\\
NR	&	New Radio\\
OS	&	Operating System\\
RAN	&	Radio Access Network\\
RU	&	Radio Unit\\
SAN	&	Stochastic Activity Network\\
SDN	&	Software-Defined Networking\\
SW	&	Software\\
UE	&	User Equipment\\
URLLC	&	Ultra-Reliable Low-Latency Communication\\
VIM	&	Virtualization Infrastructure Manager\\
VM	&	Virtual Machine\\
\bottomrule
\end{tabular}}
\label{tab0}
\end{center}
\end{table}

\section{Related Works}
\label{sec:related_works}

In literature, there are several papers~\cite{sallhammar2006towards, sallhammar2006stochastic, sallhammar2006game,9049748,NguyenmedIoT} that introduce models that jointly model security and dependability of different information systems, but none of these works address a whole 5G-MEC system.
Sallhammer et al.~\cite{sallhammar2006towards, sallhammar2006stochastic, sallhammar2006game} focus on the interplay between security and dependability in information systems. The authors propose a stochastic modeling approach that uses game theory concepts to predict and quantify attacker behavior and system vulnerabilities. They argue that this model facilitates a more realistic assessment of system security and dependability by considering accidental faults and intentional attacks within a unified framework. These works focus on generic information systems such as Domain Name System (DNS) servers and do not consider communication systems like in 5G. 
Kang et al.~\cite{9049748} introduce a model that evaluates the security and dependability of network systems under the threat of lateral movement attacks. This model is instrumental in understanding network systems' vulnerabilities and necessary defense mechanisms when multiple devices are in the same network, highlighting an integrated security and dependability approach. Although this work models devices in a network, it does not include anything regarding 5G or MEC. 
Nguyen et al. \cite{NguyenmedIoT} model the security and dependability of Internet-of-Things (IoT) medical monitoring systems, broadly discussing the edge computing platform. They assess the security and dependability of medical IoT devices, communication channels, and data storage mechanisms, emphasizing the importance of secure and reliable IoT systems in healthcare. Even though these proposed models include the concept of edge computing, they do not consider 5G or any specifics of MEC.
Liu et al.~\cite{9605723} propose a security and dependability model for MEC data centers, focusing on state transitions associated with lateral movement attacks. The authors highlight the significance of considering both security and dependability to ensure seamless operation and sustained availability. Even though they model the MEC hosts, they do not consider 5G connectivity or the rest of the MEC system.

Some works also focus on modeling 5G-MEC systems, but these models individually target security or dependability. Carvalho et al.~\cite{9144176} introduce a novel approach for managing Virtual Machines (VMs) in 5G-MEC systems, focusing on balancing security enhancement with performance optimization. The work employs a semi-Markov decision process to manage service rejection and security risks. The paper also introduces a new metric for security risk quantification and a cost structure for evaluating the impact of deploying additional VMs. This work primarily focuses on the security related to the hypervisor-related management for MEC hosts and does not consider the whole 5G-MEC system or its dependability.  

Huang et al.~\cite{HUANG2019755} present a security model for optimizing security and computational overhead in MEC environments. The model provides a framework for calculating the security overhead for various services, considering factors like dataset size and MEC server specifications, aiding in efficient security service deployment decisions. This work focuses only on the security of the MEC servers rather than the whole MEC system, with 5G as the connectivity medium.

In our previous work~\cite{pathirana2023availability}, we introduce a dependability model focusing on the availability of the whole 5G-MEC systems. We use a two-level modeling technique to consider the key elements of the system and then the different components of each element. However, we do not consider the connectivity between the elements or the security aspect of the 5G-MEC system. We discuss those models in more detail in the following sections. 

In summary, this paper has the following novelties:
\begin{itemize}
    \item It jointly models the security and dependability of a 5G-MEC system.
    \item It considers a whole 5G-MEC system including the following elements:
    \begin{itemize}
        \item 5G radio and core networks;
        \item MEC hosts and management;
        \item connectivity between all these elements.
    \end{itemize}
\end{itemize}

\section{Preliminaries} \label{sec:background}

This section presents the necessary background, including the 5G-MEC architecture, the specific implementation we further consider for our model, and some notes on attacks against the availability in 5G-MEC systems.

\subsection{5G-MEC System}

\subsubsection{General Architecture}

The 5G system primarily comprises two main components: the 5G Core (5GC) and the 5G Radio Access Network (RAN). The 5GC serves as the central part of the network, handling critical management and orchestration functionalities. It is tasked with crucial operations such as session management, where it maintains and manages the state of user connections; user authentication, which involves verifying user credentials to ensure security; and policy enforcement, which applies rules governing network use to manage resources and user access effectively. The RAN comprises 5G base stations, known as the gNodeBs, which connect devices to the 5GC by employing new technologies like massive MIMO and beamforming to enhance connectivity and bandwidth. Together, these components facilitate the high-speed, low-latency communications of 5G networks~\cite{ETSI:5G:123.501}.

As specified by the European Telecommunications Standards Institute (ETSI)~\cite{ETSI:MEC:003}, the MEC architecture is organized in two layers: the \textit{MEC Host Level} and the \textit{MEC System Level}. The key elements of the MEC Host Level are the MEC Hosts (MEHs), which are the actual edge computing devices. An essential component of the MEH is the MEC Platform (MEP), which is a collection of functionalities needed to run the various MEC applications.
The other elements of the MEC Host Level are the Virtual Infrastructure Manager (VIM) and the MEP Manager (MEPM), responsible for managing the operations and functionalities of individual MEH, such as hosting MEC applications and providing services at the network's edge. The MEC System Level mainly comprises the MEC Orchestrator (MEO), which coordinates among multiple MEHs and oversees broader operational aspects like orchestrating resources across the network and ensuring overall system efficiency and reliability. 

The integration of 5G and MEC architectures, as described by ETSI~\cite{ETSI:MEC:031}, is strategically designed to harness the combined strengths of high-speed and low-latency capabilities by 5G and local data processing by MEC.
In particular, the MEC is seen by 5G as an Application Function, and a MEC is deployed as a Data Network. The integration of 5G and MEC is performed by the User Plane Function, which is the 5GC function that supports features and capabilities to facilitate data plane operations.

\subsubsection{Specific Implementation}

\begin{figure*}[ht] 
\centering 
\includegraphics[width=14cm]{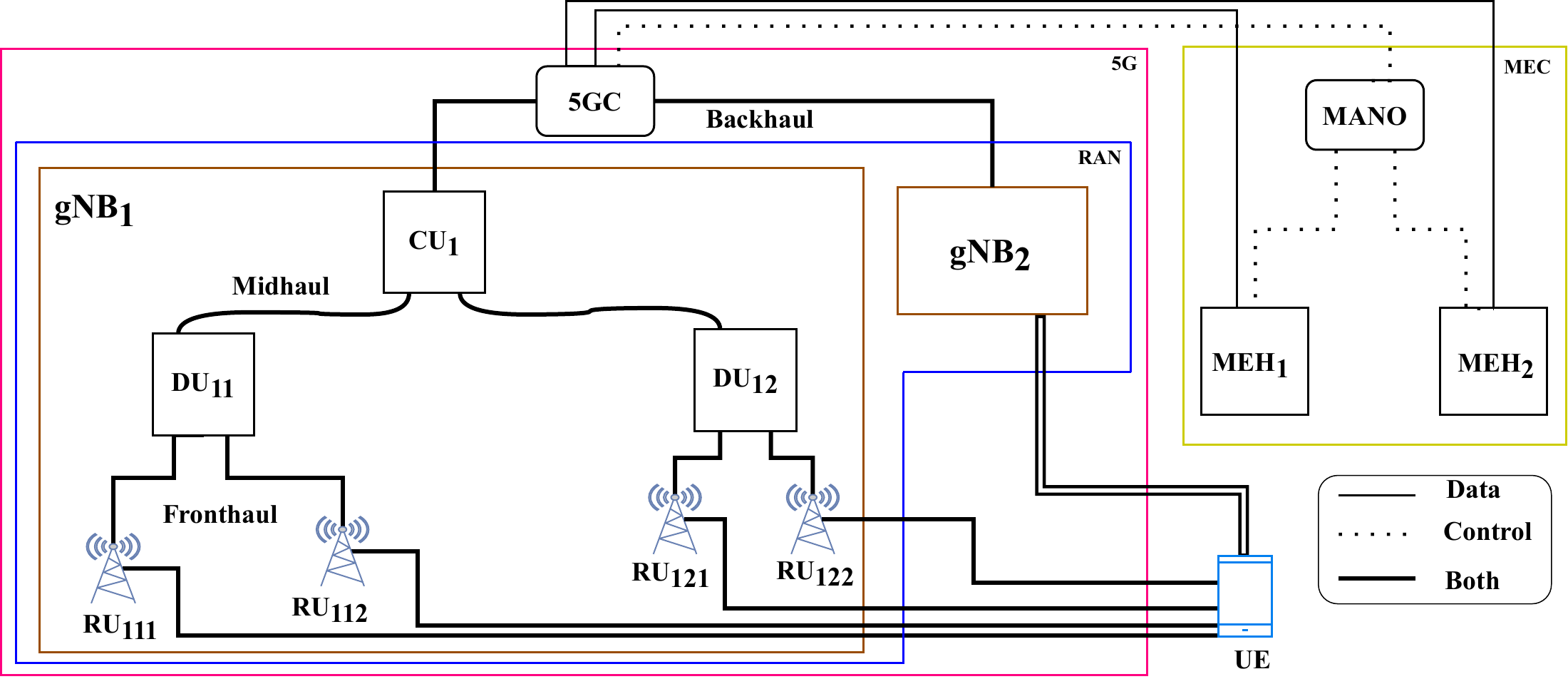}
\caption{Representation of the 5G-MEC system under investigation} 
\label{fig:scenario1} 
\end{figure*}

Figure~\ref{fig:scenario1} illustrates the structure and associated interconnections of the 5G-MEC system under investigation. 

In the 5G-MEC systems that will be modeled, the RAN is implemented as an Open RAN (ORAN)~\cite{oran_spec}. ORAN is an initiative to create more open and interoperable RAN architectures. In order to achieve this target, ORAN implements the disaggregation and decomposition of the gNodeBs. 
The disaggregation consists of the separation between the hardware and the software composing a gNodeB, which will be implemented in software running on generic hardware platforms. The disaggregation eliminates vendor lock-in and enables the selection of the best combination of hardware and software vendors.
The decomposition consists of the split of the gNodeB into multiple functions, which can be virtualized. The decomposition enables the utilization of virtualized network functions from multiple vendors, which can be deployed on Commercial-Off-The-Shelf (COTS) servers by using merchant silicon~\cite{NGMN:decomp}.

In ORAN, the traditional functions of a gNodeB are divided into three distinct units: the Central Unit (CU), the Distributed Unit (DU), and the Radio Unit (RU). Each unit has specific roles. The RU is primarily responsible for radio transmissions that communicate directly with the User Equipment (UE), such as smartphones and other devices. This involves sending and receiving radio waves, managing signal strength, and ensuring clear communication with minimal interference~\cite{ETSI:5G:138.401}.

The DU has the task of processing data received from the RU and forwarding it to the core network. It acts as a middleman that transmits data and performs initial processing, which may include error checking and some data optimization to ensure efficient transmission over the network. Multiple RUs can connect to a single DU, facilitating scalability and flexibility in network management~\cite{ETSI:5G:138.401}.

The CU oversees several DUs, providing higher control and resource orchestration~\cite{ETSI:5G:138.401}. It plays a paramount role in the overall efficiency and performance of the network by managing the distribution of network resources among the DUs. This includes optimizing data routes to reduce latency and improve speed, which is essential for maintaining the high performance expected of 5G networks. The enhanced capabilities of the CU in orchestrating network resources are vital for supporting advanced 5G features such as URLLC and massive machine-type communications, which are crucial for applications ranging from industrial automation to critical emergency services.

ORAN's approach to disaggregate and decompose in network functions and to utilize open interfaces potentially enlarges the attack surface, especially in a public MEC environment where resources and services are shared across various users and operators. This openness makes it easier for potential attackers to exploit vulnerabilities. It complicates securing the multiple interfaces against inconsistencies in security practices among different vendors~\cite{zeyu2020survey}. Furthermore, integrating various components from different suppliers might result in compatibility issues and performance inconsistencies. These are particularly problematic in environments that demand high reliability and low latency, such as those required for autonomous driving or real-time data analytics~\cite{foukas2017network}.

In terms of dependability, the modular nature of ORAN components sourced from a diverse set of vendors might lead to operational misalignments, increasing the likelihood of system failures or downtimes. This is a critical concern in public MEC setups where continuous uptime and consistent network behavior are essential for safety-critical applications~\cite{tran2017collaborative}. 
Moreover, the integration of ORAN with the MEC infrastructure, particularly within a public MEC setting, introduces complex security, performance, and dependability challenges~\cite{checko2014cloud, garcia2018fluidran}.

In our model, we consider an ORAN implementation for two reasons:
(i) as presented, ORAN is challenging from both security and dependability perspectives;
(ii) network operators are building ORAN networks because it significantly enhances network flexibility, efficiency, and innovation through open interfaces and standardized components \cite{Vodafone}. 

RU, DU, and CU are implemented as discussed in \cite{telefonica,telecominfra}.
The RU is a specialized device composed of dedicated hardware (HW), Firmware (FW), and antennas.
DU and CU are both implemented on COTS HW, which runs an Operating System (OS), and the Software (SW), which implements the DU and CU functionalities.
In alignment with the ORAN specification and given the different levels of criticality, the CU HW is redundant. Instead, the DU HW is not.
In particular, we consider a 1+1 active standby HW redundancy for the CU.

We assume that the various functions of the 5GC are running on top of a generic OS in a centralized environment; the comprehensive review \cite{9670414} provides further details of similar deployments of 5GC. We selected this implementation because of its simplicity but also because it is a challenging scenario from dependability and security perspectives.

For the simplicity of modeling, we also consider the MEO, MEPM, and VIM as a single management-and-orchestration (\textit{MANO}) element. This approach has also been used for modeling the MANO of the Network Function Virtualization (NFV)~\cite{tola2019network}. NFV is an architecture for the virtualization of network services by using virtual network functions~\cite{ETSI:NFV:006}. ETSI has also defined an implementation of MEC by using NFV~\cite{ETSI:MEC:003}. We consider a setup with SW, OS, and HW that is mainly based on Open Source MANO (OSM), which is a platform for the orchestration of NFV \cite{9324847}.

We assume that both 5GC and MEC MANO are implemented in a cluster. The cluster is composed of multiple SW and HW instances that share the computational load and enable redundancy. This approach is similar to the one used to model the Software-Defined Networking (SDN) controller~\cite{nencioni2016dsn,nencioni2017including}. The SDN is a networking paradigm that separates the control plane and the data plane and logically centralizes the control plane in one element, the controller~\cite{nencioni2018orchestration}. Therefore, the SDN controller is a critical management element whose implementation is assumed to be similar to that of 5GC and MEC MANO.

The MEHs are implemented using virtualization as defined by ETSI~\cite{ETSI:MEC:027}.
As in the reference paper~\cite{pathirana2023availability}, we consider an MEH to be a type-II virtualization-enabled COTS device, where a host OS runs the MEP and the MEC application on top of VMs using a hypervisor
\cite{lee2019case}.

Finally, we present the connections between the elements of the 5G-MEC system. \textit{Backhaul} is the connection between the gNodeB and the core, \textit{midhaul} between CU and DUs when they are located at different physical locations, and \textit{fronthaul} between DUs and RUs within a gNodeB. The connections in the backhaul, midhaul, and fronthaul can be either a single physical link or a path in a network, and they are used by both the control and the data planes. The \emph{air interface} is the connection between RUs and UE, is a single wireless link, and is used by both the control and the data planes. The connections between the MANO and 5GC and the MANO and MEHs are control connections, while the connections between 5GC and MEHs are data connections (they are part of the path between the UE and the MEHs). As mentioned earlier, ETSI defines multiple deployment methodologies for 5G-MEC systems~\cite{ETSI:MEC:031}. Each deployment method might feature a distinct system architecture and varying interconnections between components. For all the connections except the air interface, we assume they are a path in a network since it is the most challenging scenario from both dependability and security perspectives.

Even if tailored for this implementation of the 5G-MEC systems, the proposed model can adapted to represent other implementations.

\subsection{Attacks on 5G-MEC Systems}
\label{subsec:5gmec_attacks}

In the context of a 5G-MEC system, where low latency and high reliability are essential, and the provided services are many times critical, Denial-of-Service (DoS) (and, of course, the more powerful variation Distributed DoS (DDoS)) poses a significant threat.
DoS attacks can lead to minor disruptions or major falls. They might serve as diversions for graver threats like data breaches or malware deployment~\cite{gu2007denial} or can lead to hardware or software failures because of overheating caused by overloaded components or buffer overflow~\cite{6825828, buffeover}. Moreover, the use of not component-centric equipment such as COTS (heavily used in OpenRAN) and virtualization for 5G deployments has opened the path to many DoS attack scenarios~\cite{ettiane2021toward}. As a consequence, our attack modeling spotlights DoS attacks as a representative threat for all 5G-MEC elements~\cite{8792139}. 

A typical DoS attack involves jamming the RU and the UE connection. This type of attack is particularly concerning in the context of 5G networks due to their reliance on high-speed, low-latency communications. Research in wireless network security has extensively documented the impact of jamming and other DoS attacks on network components and connections~\cite{5473884, birutis2022study, 9031175, RUjam}. These attacks can degrade the performance, reliability, and availability of network services, posing significant threats to the functionality of wireless communication systems.

\section{Dependability and Security Modeling}
\label{sec:dep_sec_modeling}

This section refers to joint dependability and security modeling, basic concepts required to understand our proposed model, and the dependability model \cite{pathirana2023availability} we use as a primary reference and security modeling techniques we use as inspiration for our work.

\subsection{Joint modeling}

The modeling of dependability is usually different from the modeling of security. While dependability models are primarily quantitative in the sense that they aim to compute values for attributes of interest  (e.g., the percentage of time the system is available), security models often target the success probability of attacks that aim to break such attributes (e.g., the success probability of a well-defined adversary to break availability). Security models are more complex in the sense that they need to model the adversary, too (e.g., abilities, behavior, time, and computational resources).

A joint model must accommodate both needs, which is demanding. Our work follows the line of previous papers~\cite{pathirana2023availability,NguyenmedIoT,9605723, 9049748} and considers a two-layer approach that makes use of commonly used techniques for the quantitative evaluation of the availability of a 5G-MEC system. More precisely, we construct our models starting from \cite{pathirana2023availability}, which makes use of a Fault Tree (FT) and Stochastic Activity Networks (SANs). For completeness, we briefly explain FTs and SANs next. For more details, the reader can refer to~\cite{ericson1999fault, meyer1985stochastic}.

\subsubsection{FT}

FT is an extensively used method in assessing system reliability, maintainability, and safety. FT systematically represents the various parallel and sequential combinations of faults that could result in system failures. FT illustrates how subsystem and element failures propagate through a system by depicting potential faults as a tree rooted in the undesired event, i.e., failure of the whole system. This visual representation includes logic gates (mostly OR and AND gates) to show the relationship between events, allowing engineers to qualitatively and quantitatively analyze the probability of the primary failure. This thorough analysis aids in identifying critical components, guiding preventive maintenance, and system design improvements~\cite{ericson1999fault}.

\subsubsection{SAN}

A SAN is an advanced modeling tool that analyzes complex systems with dynamic activities and randomness over time, extending the Stochastic Petri Net (SPN). In SAN models, the system is depicted as a network comprising places (representing states or conditions) and activities (representing the transitions between places)~\cite{meyer1985stochastic}. A place may contain tokens, and the number of tokens in each place (marking) identifies the system's current state. The activities may be characterized by rates or probability distributions related to the movement of tokens from one place to another. To model diverse behaviors, SAN models include both timed activities occurring after specific time intervals and immediate activities happening instantaneously. An activity may have two or more cases with associated probabilities, which are used to select the succeeding place after the completion of the activity.
Moreover, SAN models include gate functions, which act as conditions for activity occurrence or force complex token firing once an activity occurs. 

\subsection{Reference Dependability Model}
\label{subsec:dep_model}

We base our modeling on \cite{pathirana2023availability}. The model adopts a two-tiered approach, which we maintain in our proposal. 

\begin{figure*}[t!]
    \centering
    \includegraphics[width=10cm]{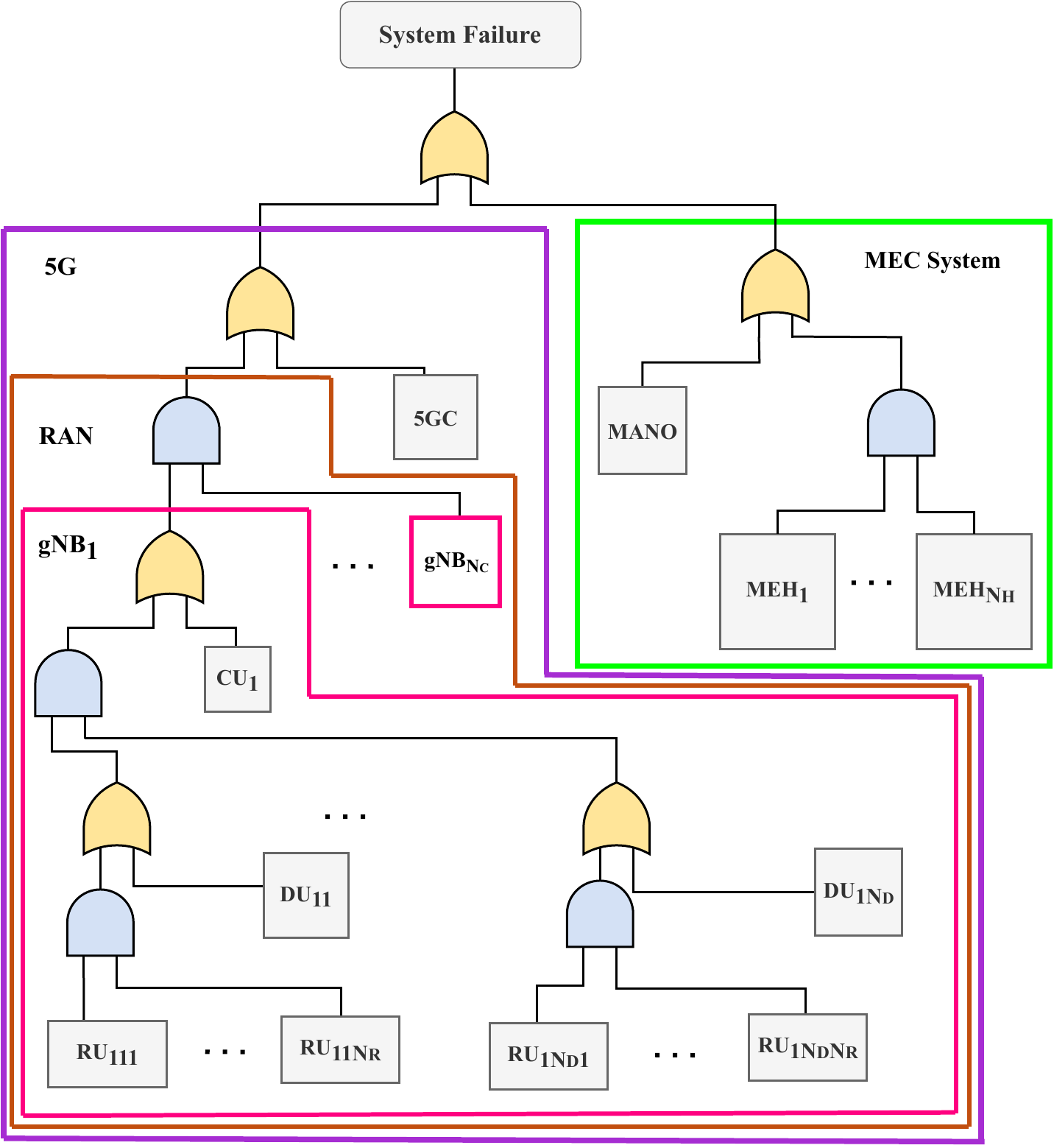}
    \caption{ FT model of the 5G-MEC system~\cite{pathirana2023availability}}
    \label{fig:FTL1}
\end{figure*}

\begin{table}
    \caption{Notations}
\centering
\begin{tabular}{p{0.15\linewidth} p{0.65\linewidth}}
\toprule
Notation &Description \\ 
\midrule

$N_R$        & Number of redundant RUs per DU\\
$N_D$        & Number of redundant DUs per CU\\
$N_C$        & Number of redundant CUs \\
$N_H$        & Number of redundant MEHs \\

\bottomrule
\end{tabular}
    \label{table:notations}
\end{table}

The first level represents the interrelationships among the elements of the 5G-MEC system in order to determine the overall unavailability of the system. This first level, illustrated in Figure~\ref{fig:FTL1}, expresses the system unavailability $U_{sys}$ in terms of the unavailabilities of the individual elements RU, DU, CU, 5GC, MANO, and MEH by using the Boolean logic. Table~\ref{table:notations} explains the notations in terms of redundant elements. For example, the 5G-MEC system is unavailable if the 5G or the MEC system is unavailable (OR gate), and the MEC system is unavailable if the MANO or the MEHs (OR gate) are all down (AND gate). Overall, from \cite{pathirana2023availability}:

\begin{equation*}
\resizebox{.97\hsize}{!}{$
U_{Sys} = 1 - \left[\left(1 - U_{RAN}\right)\left(1 - U_{5GC}\right)\left(1 - U_{MANO}\right)\left(1 -  U_{MEH}^{N_H}\right)\right]
$},  
\end{equation*}
where
\begin{equation*}
\resizebox{.97\hsize}{!}{$
U_{RAN} = \left[1 - \left(1 - \left(1 - \left(1 - U_{RU}^{N_R}\right)\left(1 - U_{DU}\right)\right)^{N_D}\right)\left(1 - U_{CU}\right)\right]^{N_C}
$}.
\end{equation*}

This is a general way to express the system unavailability in the given architecture, so we keep this level unchanged.

The second level utilizes SANs to model each element's dependability. We again start from \cite{pathirana2023availability} but modify this level accordingly to accommodate security and connectivity issues between the system elements. 

\subsection{Security Modeling}

We further mention some specific models that inspired our security extension. They have been already introduced in Section~\ref{sec:related_works}, we now provide more details on these models.

\begin{figure}[t!]
    \centering
    \includegraphics[width=8cm]{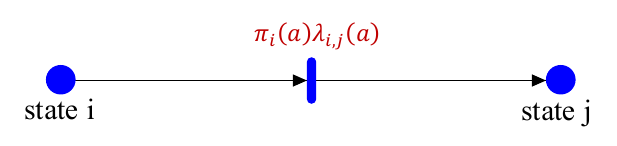}
    \caption{SAN security model (adapted from \cite{sallhammar2006towards, sallhammar2006stochastic, sallhammar2006game})} 
    \label{fig:secmod}
\end{figure}

Sallhammar et al. \cite{sallhammar2006towards, sallhammar2006stochastic, sallhammar2006game} model security as follows. An adversary can choose to perform an action that exploits an existing vulnerability with the aim to harm the availability of a system. Let $\pi_i(a)$ be the probability of the adversary choosing a malicious action $a$ when the system is in the state $i$. Let $\lambda_{i,j}(a)$ be the expected time to exploit the vulnerability and thus place the system from a properly functioning state $i$ to a security failed state $j$. Then the failure rate from $i$ to $j$ is given by $\pi_i(a)\cdot\lambda_{i,j}(a)$. Figure~\ref{fig:secmod} illustrates the corresponding model.

Nguyen et al. \cite{NguyenmedIoT} model components as Continuous Time Markov Chain (CTMC). The authors consider fine-grained security modeling, including several aspects such as detection of, adaptation to, and recovery from attacks. The model considers several states between the initially healthy state $N$ and the failure state $F$. These include a vulnerable state $V$ (the adversary had performed some actions and identified existing vulnerabilities that can further be exploited), a compromised state $C$ (the adversary had performed some compromising actions), an under-attack state $A$ (the system is under attack), an adaptation state $AD$ (the system tries to adapt to the detected attacks), and a rejuvenation state $R$ (the system removes existing vulnerabilities, restarts, or perform other clearing actions). The transition from one state to another is characterized by a meantime. At the same time, the success of an action (e.g., the success or failure of the rejuvenation) is represented by a coverage factor. 

We will explain how we inspired from these models while presenting our model extensions in the next section.
\section{Security-aware Availability Model} \label{sec:model}

As already mentioned, we incorporate all the changes in the second level, the SAN models. The SAN models now incorporate attacks against the 5G-MEC elements, which might lead to the unavailability of the system. These attacks can impact an element itself (seen as stand-alone) or its connectivity with other elements. Similarly, the SAN models also incorporate the failure of the connectivity between elements because of non-intentional events. 
To avoid the duplication of a connection (i.e., consider the failure of a connection twice), we organize the models as follows. In general, we include the connectivity between two elements in the model of the lower element (i.e., the element closer to the user): the RU-DU connection in the RU model, the DU-CU connection in the DU model, the MEH-MANO connection in the MEH model, etc. As we are not interested in modeling the UE, we model the connection between the UE and a RU within the RU model. Finally, we choose the connection between the MANO and the 5GC within the model of the MANO. Same as \cite{pathirana2023availability}, our model assumes uncorrelated, independent failures. Looking into correlated failures is the target of future work.

\begin{table}
   \caption{Default intensity values}
\begin{center}
\adjustbox{width=0.4\textwidth,totalheight=0.95\textheight,keepaspectratio}{
\begin{tabular}{p{0.15\linewidth} p{0.21\linewidth} p{0.61\linewidth}}
\toprule
Intensity &Value &Description [Mean time to]\\ 
\midrule

$1/\lambda_{RH}$        & 17 years                  & RU HW failure         \\
$1/\mu_{RH}$            & 6 hours                   & RU HW repair/recovery       \\
$1/\lambda_{HW}$          & 6 months                  & HW failure                \\
$1/\mu_{cov}$           & 30 minutes                & manual coverage               \\
$1/\mu_{HW}$            & 2 hours                   & HW repair                \\
$1/\mu_{HW_{fo}}$       & 3 minutes                 & HW failover             \\
$1/\lambda_A$           & 104 months                & antenna failure          \\
$1/\mu_{A}$             & 6 hours                   & antenna repair/recovery         \\
$1/\lambda_{FW}$        & 75 days                   & FW failure            \\
$1/\mu_{FW}$            & 65 minutes                & FW repair/recovery          \\
$1/\lambda_{OS}$        & 2 months                  & OS failure           \\
$1/\mu_{OS}$            & 1 hour                    & OS repair          \\
$1/\mu_{OS_{r}}$        & 1 minute                  & OS reboot          \\
$1/\mu_{HYP_{rs}}$      & 2.5 minutes               & restart of hypervisor and VMs          \\
$1/\lambda_{HYP}$       & 4 months                  & hypervisor failure              \\
$1/\mu_{HYP}$           & 1 hour                    & hypervisor repair            \\
$1/\mu_{HYP_{r}}$       & 1 minute                  & hypervisor restart             \\
$1/\mu_{VM_{rs}}$       & 1.5 minute                & restart of VMs          \\
$1/\lambda_{VM}$        & 3 months                  & VM failure  \\
$1/\mu_{VM}$            & 1 hour                    & VM repair  \\
$1/\mu_{VM_{r}}$        & 1 minute                  & VM reboot              \\
$1/\lambda_{APP}$       & 2 weeks                   & application failure             \\
$1/\mu_{APP}$           & 30 minutes                & application repair             \\
$1/\mu_{APP_{r}}$       & 15 seconds                & application software restart              \\
$1/\lambda_{SW}$        & 1 month                   & SW failure           \\
$1/\mu_{SW}$            & 30 minutes                & SW repair           \\
$1/\mu_{SW_{r}}$        & 30 seconds                & SW restart        \\
$C_{HW}$                & 0.97                      & coverage factor for HW failover        \\
$C_{OS}$                & 0.9                       & coverage factor for OS reboot/failover              \\
$C_{HYP}$               & 0.9                       & coverage factor for hypervisor restart              \\
$C_{SW}$                & 0.85                      & coverage factor for SW restart/failover             \\
$C_{VM}$                & 0.9                       & coverage factor for VM reboot            \\
$C_{APP}$               & 0.8                      & coverage factor for APP restart              \\
$(M,K)$                 & (10,8)                   & cluster settings             \\
\hline
$1/\lambda_{R}$ & 15 days & DoS attack \\ 
$1/\mu_{R}$ & 30 minutes & DoS recovery \\ 
$1/\lambda_{S}$ & 60 minutes & successful attack\\
$1/\mu_{S}$ & 5 minutes & attack detection\\
$1/\lambda_{T}$ & 2 months & attack of the management components\\
$1/\mu_{T}$ & 35 minutes & recovery from management components' attack\\
$1/\lambda_{P}$ & 60 minutes & successful attack during detection\\
$1/\mu_{P}$ & 10 minutes & automatic adaptation \\ 
$1/\lambda_{M}$ & 60 minutes & successful attack during manual adaption\\
$1/\mu_{M}$ & 60 minutes & manual adaption \\ 
$1/\lambda_{U}$ & 15 days & UE-RU connection failure \\
$1/\mu_{U}$ & 15 minutes & UE-RU connection recovery \\
$1/\lambda_{C}$ & 4 months & generic connection failure\\
$1/\mu_{C}$ & 15 minutes & generic connection recovery \\
$C_{P}$ & 0.8 & probability of a successful adaption \\
$C_{D}$ & 0.9 & probability of a successful attack detection\\
$C_{U}$ & 0.8 & probability of a successful attack on RU\\
$C_{S}$ & 0.9 & probability of a generic successful attack\\
$\gamma$ & 10 & failure amplification factor\\
\bottomrule
\end{tabular}}
    \label{table:values}
\end{center}
\end{table}

\subsection{RU Model}
\begin{figure}[t!]
\centering
    \includegraphics[width=\columnwidth]{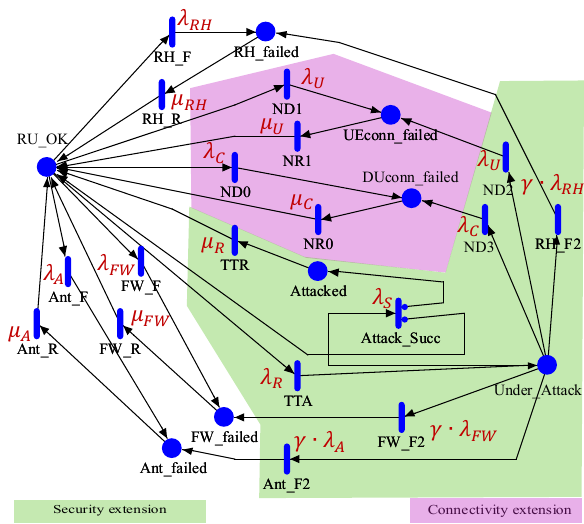}
    \caption{SAN model of a RU}
    \label{fig:SAN:RU}
\end{figure}

Figure~\ref{fig:SAN:RU} depicts the SAN model of a RU and its connections with the UE and the associated DU. The green area highlights the proposed security extension, and the purple area highlights the proposed connectivity extension. The figure includes the rates of each timed activity, where the values and definitions of the rates are listed in Table~\ref{table:values}.

The token is initially placed in \emph{RU\_OK}. As in \cite{pathirana2023availability}, we consider that the failure of each of the RU components (RU HW (RH), FW, and set of antennas (Ant)) leads to the failure of RU (\emph{RH\_failed}, \emph{FW\_failed}, and \emph{Ant\_failed}, respectively). To explain the notations, the rates $\lambda$ correspond to activities that place the token in failed states, while the rates $\mu$ denote recovery rates that correspond to activities that place the token back to \emph{RU\_OK}.

The \textbf{connectivity extension} considers the failure of the two connections UE-RU and RU-DU, which lead to \emph{UEconn\_failed}, and \emph{DUconn\_failed}, respectively. Again, each connectivity failure is characterized by a corresponding failing rate ($\lambda_U$ and $\lambda_C$), as well as a corresponding recovery rate ($\mu_U$ and $\mu_C$, respectively). The two connectivity failures are unintentional (i.e., not triggered by an adversary) and might happen either when the RU is functioning normally (\emph{RU\_OK}) or when the RU is under attack (\emph{Under\_Attack}). We ignore connectivity failures directly caused by attacks because we assume they are contained within a successful attack and hence included in the security extension.

The \textbf{security extension} considers an attack against the RU or its connections (UE-RU and RU-DU). We intentionally do not differentiate an attack against the element itself from an attack against its connections for several reasons, including the simplicity of the model exposure and the strong relation between the two (an adversary succeeds in breaking availability regardless of whether the adversary puts down the element or its connectivity within the system). We model the RU as being under attack (\emph{Under\_Attack}) from the time the adversary mounts the attack until the adversary either succeeds (\emph{Attacked}) or fails. In our model, an unsuccessful attack is an attack that returns to the fully functional state \emph{RU\_OK}. Following a common approach in security evaluation, the success of the attack is given by a probability $C_U$.

The way we model security differs from the modeling in \cite{sallhammar2006towards, sallhammar2006stochastic, sallhammar2006game} in two ways: (i) we assume the adversary will always attack (i.e., the probability $\pi_i(a)$ to start an attack equals 1) and (ii) we consider the attack will succeed with a given probability ($C_U$). Similar to \cite{sallhammar2006towards}, and under the same motivation, we maintain the negatively exponential distribution of the expected time to succeed or fail an attack ($\lambda_S$); moreover, we consider the same type of distribution for the time to start the attack ($\lambda_R$). The motivation lies in the simplification of the assessment of the model, as well as in maintaining a similitude to the dependability model. As an addition over the models in \cite{sallhammar2006towards, sallhammar2006stochastic, sallhammar2006game}, and inspired by \cite{NguyenmedIoT}, we include the \emph{Under\_Attack} place, reachable when the adversary starts to perform compromising actions that might cause the failure of the system. The existence of such a place, as well as the detection- and adaptation-related places for the rest of the components excluding the RU (also inspired by \cite{NguyenmedIoT}), allows more fine-grained modeling.
As a difference from \cite{NguyenmedIoT}, to express the success of an action, we use conditional transactions and associate some probability of success (e.g., $C_U$, the probability of a generic successful attack). Using probabilities to evaluate the adversarial success is a common, natural practice in security evaluation.

The RU is unavailable if there is a token in any of the failed or attacked places: \emph{RH\_failed}, \emph{FW\_failed}, \emph{Ant\_failed}, \emph{UEconn\_failed}, \emph{DUconn\_failed}, or \emph{Attacked}.
Note that the element's components or connections can fail at any time, including the period when the element is under attack. Hence, we consider transactions from \emph{Under\_Attack} place to any other failed place.
Moreover, when the RU is under attack, its components are under stress, leading to an increased likelihood to fail~\cite{6168428,cherdantseva2016review}. We have modeled this effect by using an amplification factor $\gamma$ to the failure rate of each RU's component (HW, FW, and antenna).

As a difference from the next components (starting with the DU), we ignore attack detection for the RU. This is because the number of RUs in the network is high, and the implementors may consider deploying cheaper products to maximize profit, which might not ship with automatic attack detection or recovery~\cite{LIYANAGE2023103621}.  
After a successful attack, the RU can go back to its normal functioning state \emph{RU\_OK}, as a result of a manual process (e.g., manual reset, hardware or software fix). 

\subsection{DU Model}

\begin{figure}[t!]
    \centering
    \includegraphics[width=\columnwidth]{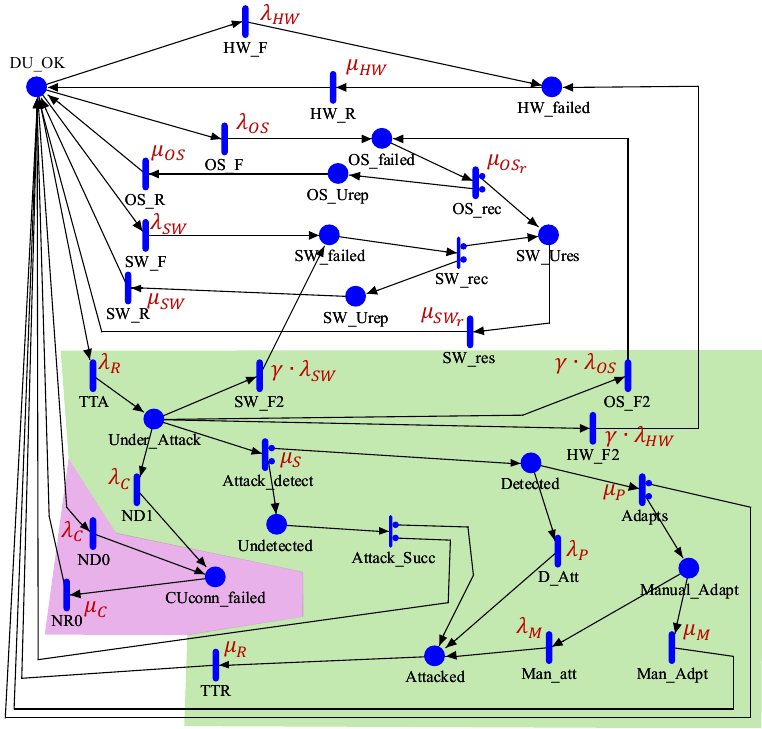}
    \caption{SAN model of a DU}
    \label{fig:SAN:DU}
\end{figure}

Figure~\ref{fig:SAN:DU} depicts the SAN model of a DU and the connection with its corresponding CU. As with the RU model, the connectivity and security extensions are marked in green and purple, respectively. The values and descriptions of the rates of each timed activity are listed in Table~\ref{table:values}.

As mentioned in Section~\ref{sec:background}, the DU has three components (HW, OS, and SW)
whose individual failures lead to \emph{HW\_failed}, \emph{OS\_failed}, and \emph{SW\_failed}, respectively. In case of an OS failure, the OS reboots. With probability $C_{OS}$, the reboot succeeds, and then the SW restarts (\textit{SW\_Ures}). Otherwise, the hard repair is necessary (\textit{OS\_Urep}).
After an SW failure, with probability $C_{SW}$, a restart suffices to bring the element back to the functional state. Otherwise, the hard repair is necessary (\textit{SW\_Urep}).

The \textbf{connectivity extension} models the failure of the CU-DU connection (\emph{CUconn\_failed}) similarly to the RU model: the connection to the CU can fail with a failing rate $\lambda_C$ and a corresponding recovery rate $\mu_C$.

Compared to the RU model, the \textbf{security extension} for the DU is more complex because it includes automatic attack detection and recovery (inspired by~\cite{NguyenmedIoT}).
When the DU is under attack, the attack might be detected with a probability $C_D$. If the detection succeeds (\emph{Detected}), the DU will undergo an automatic adaptation, which succeeds with a probability $C_P$. In this case, the DU goes back to the fully functional state (\emph{DU\_OK}). Otherwise, a manual adaptation (\emph{Manual\_Adapt}) occurs.
Naturally, the attack can succeed (\emph{Attacked}) while performing the automatic (\emph{Detected}) or the manual adaptation (\emph{Manual\_Adapt}) or as a result of a failure to adapt.
In the case of an undetected attack (\emph{Undetected}), the attack can succeed (\emph{Attacked}) with a given probability $C_S$. If so, the DU will return to the fully functional state (\emph{DU\_OK}) after recovery (with an associated recovery rate $\mu_R$). Otherwise, we assume that an undetected but unsuccessful attack returns the token instantly to the \emph{DU\_OK} place.
Note that our model allows continuous detection until the attack stops (with or without success). This is because the activities originating from the \emph{Undetected} place are instant. 
 
As already mentioned for the RU, an attack can place stress on the system components, potentially leading to HW, OS, or SW failures. Specifically, a DoS attack on the DU can inundate it with overwhelming requests or data, causing it to exhaust its resources. This exhaustion could lead to system slowdowns, crashes, or even HW malfunctions if components overheat or are pushed beyond their operational limits. The OS might crash, requiring a reboot, while the SW could experience glitches, errors, or unexpected shutdowns. For this reason, when the DU is under attack (\emph{Under\_Attack}), its components can fail with a rate $\gamma$-times higher.

We neglect the HW, OS, and SW failure during the automatic adaptation (\emph{Detected}) and the manual adaptation (\emph{Manual\_Adapt}) because the average stay time is more than an order of magnitude lower than the average time to failure.

\subsection{CU Model}

\begin{figure}[t!]
    \centering
    \includegraphics[width=0.95\columnwidth]{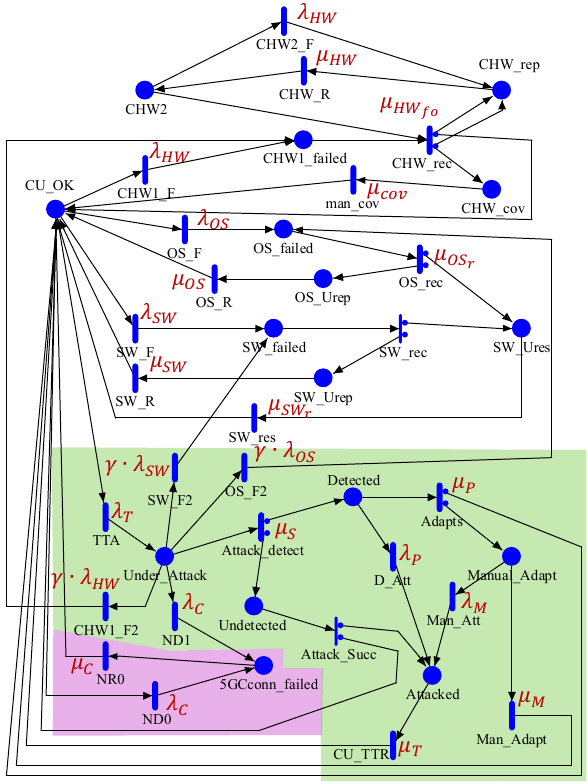}
    \caption{SAN model of a CU} 
    \label{fig:SAN:CU}
\end{figure}

Figure~\ref{fig:SAN:CU} depicts the SAN model of a CU and the CU-5GC connection. Again, the extensions are marked in color, and the necessary information is listed in Table~\ref{table:values}.

As mentioned in Section~\ref{sec:background}, the CU is configured with a 1+1 HW redundancy. The two tokens are initially set in \emph{CU\_OK} and in \emph{CHW2}.
If the primary HW fails (\emph{CHW1\_failed}) and the backup HW is working (\emph{CHW2}), then with probability $C_{HW}$ the failover is successful, the failed HW will be under repair (\emph{CHW\_rep}), and the backup HW becomes primary (\emph{CU\_OK}). Otherwise, the failover is not successful, the failed HW will be under repair (\emph{CHW\_rep}), and the backup HW needs to be fixed (\emph{CHW\_cov}) before becoming primary (\emph{CU\_OK}). The OS and SW failures are modeled as in the DU model.

Similarly to RU and DU models, the \textbf{connectivity extension} models the failure of the backhaul connection between CU and 5GC (\emph{5GCconn\_failed}).

The \textbf{security extension} follows the same ideas as for the DU. Given that the hardware is configured with redundancy, we posit that DoS attacks might lead to diminished impact on hardware. Nevertheless, note that the attack we model is generic and can include attacks against both hardware redundant components.

\subsection{MEH}

\begin{figure*}[t!]
    \centering
    \includegraphics[width=12cm]{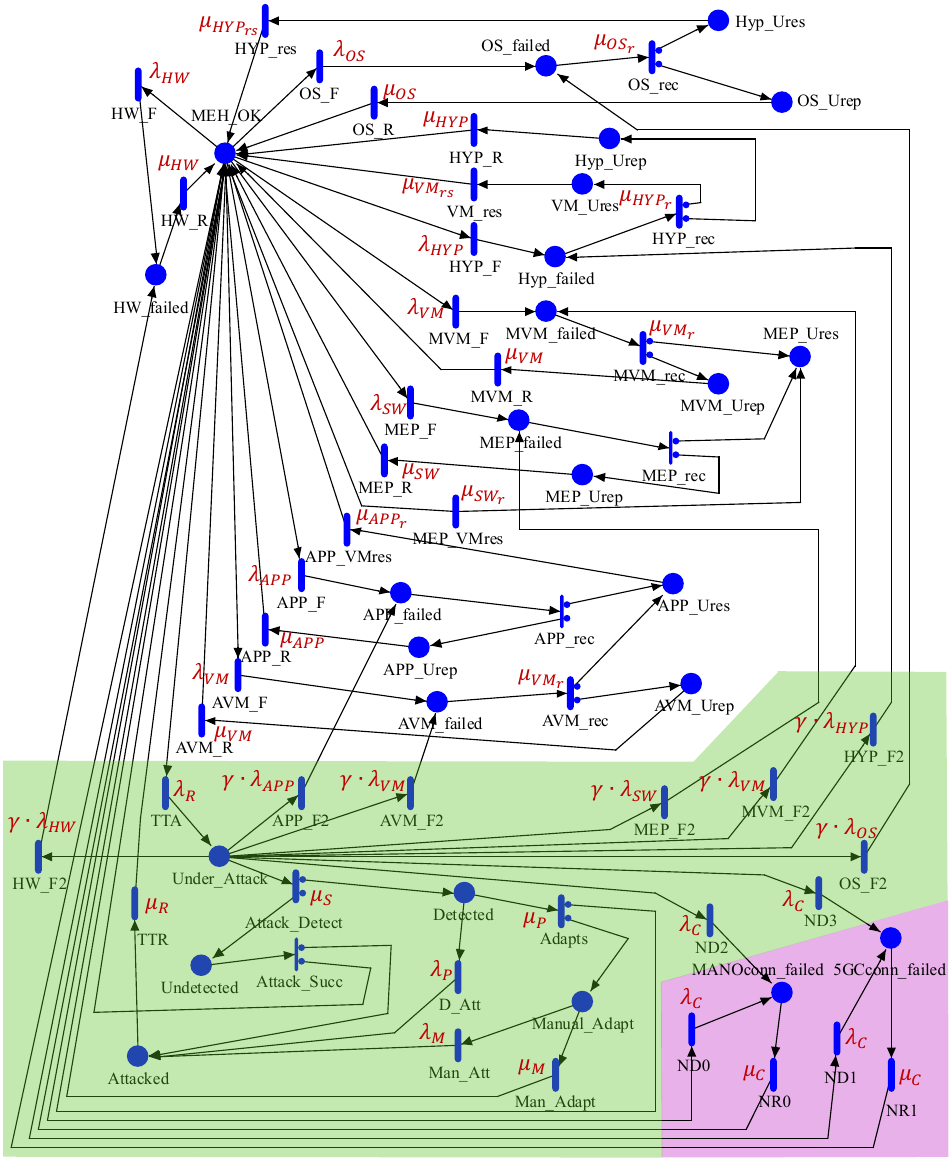}
    \caption{SAN model of a MEH} 
    \label{fig:SAN:MEH}
\end{figure*}

Figure~\ref{fig:SAN:MEH} depicts the SAN model of MEH and its connections with the 5GC and the MANO.

As explained in Section~\ref{sec:background}, we consider that an MEH is a type-II virtualization-enabled COTS device. 
On top of the shared OS, a hypervisor (HYP) runs and enables the deployment of two VMs: one runs the MEC application (APP) used by the UE, the other one runs the MEP. In case OS, HYP, MEP, APP, or VMs fail, a restart/reboot is tried first. Otherwise, a hard repair is performed. The modeling approach is similar to the one used for the DU and CU.

Similarly to the previous models, our \textbf{connectivity extension} models the connection between MEH and 5GC (\emph{5GCconn\_failed}) and between MEH and MANO (\emph{MANOconn\_failed}), and our \textbf{security extension} models the attacks in a similar way as for the DU and CU.

\subsection{5GC/MANO}

\begin{figure*}[t!]
    \centering
    \includegraphics[width=14cm]{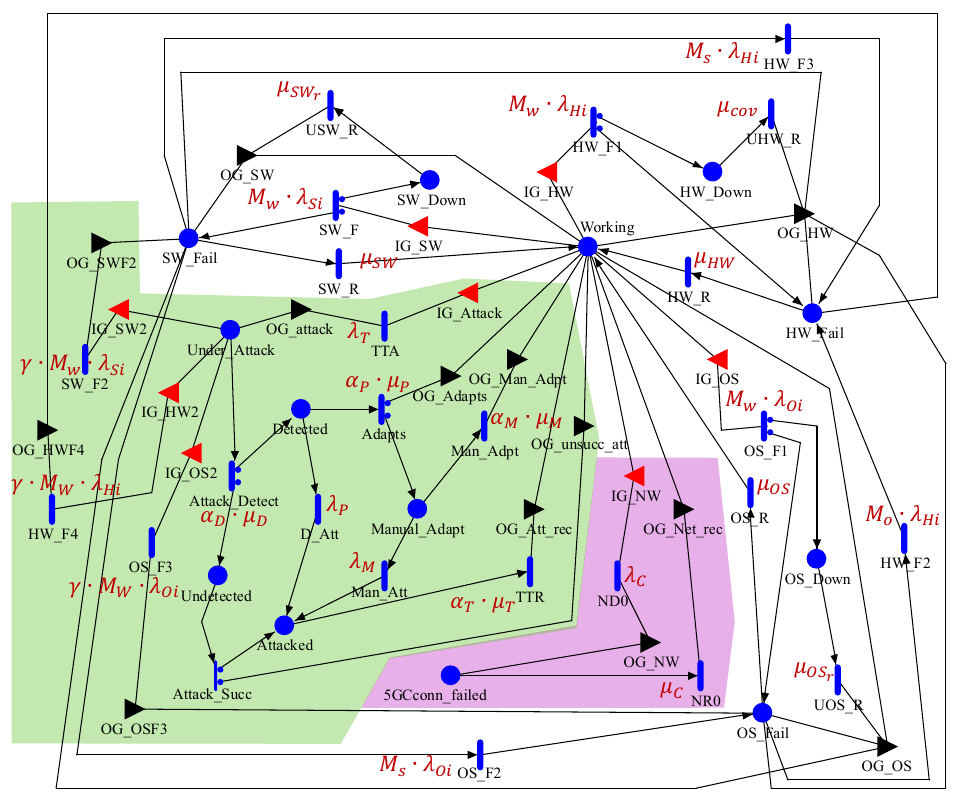}
    \caption{SAN model of a MANO} 
    \label{fig:SAN:MANO}
\end{figure*}

Following the idea in~\cite{pathirana2023availability}, we model 5GC and MANO in the same way due to the similarity of their implementations. Figure~\ref{fig:SAN:MANO} depicts the SAN model of the MANO and the connection with the 5GC. The 5GC model is similar but misses the 5GC-MANO connection. Both 5GC and MANO run in a data-center environment, which is modeled as a cluster of $M$ instances, out of which $K$ must be working for the component to work. The functions of 5GC and MANO are running as SW in the cluster~\cite{9670414,7931566}. 

As in~\cite{pathirana2023availability}, the number $M_w$ of tokens in \textit{Working} is initially set to $M$, the total number of instances in the cluster. Each instance can fail because of HW, OS, or SW. With probability $C_{HW}$, $C_{OS}$, or $C_{SW}$, the failure is limited to a single instance (\textit{HW\_Fail}, \textit{OS\_Fail}, or \textit{SW\_Fail}). Otherwise, the failure on one instance is followed by an unsuccessful failover that causes the \emph{crash} of all instances of the element (\textit{HW\_Down}, \textit{OS\_Down}, or \textit{SW\_Down}). In the mentioned context, a crash refers to a situation where all instances within a cluster fail simultaneously due to a cascading effect from a single instance failure. Such a crash might occur following the failure of a specific HW, OS, or SW instance, which leads to the operational shutdown of the entire system or cluster of instances, thus significantly amplifying the impact of the initial failure.

The input gates \textit{IG\_HW}, \textit{IG\_OS}, or \textit{IG\_SW} allow the failures to happen only when the MANO/5GC is not crashed.
We have modified these input gates to also not allow the failure to happen (by using this branch of the model) when 5GC/MANO is under attack. The failures in this case are modeled in the attack part, where the failure rate will be higher due to the stress of the attack.
The output gates \textit{OG\_HW}, \textit{OG\_OS}, or \textit{OG\_SW} allow all the failed OS and SW instances to be recovered after a crash caused by HW and OS. A SW failure causes SW instances to crash.

Similarly to previous models, the \textbf{connectivity extension} models (only for the MANO) the failure of the connection between the 5GC and the MANO (\emph{5GCconn\_failed} place).
The input gate \textit{IG\_NW} allows a failure only if the connection has not already failed. The output gates \textit{OG\_NW} and \textit{OG\_net\_rec} add and delete a token in \emph{5GCconn\_failed}, respectively, without modifying $M_w$.

The \textbf{security extension} models the attacks in a similar way to the DU, CU, and MEH. The input gate \textit{IG\_Attack} allows an attack only when the 5GC/MANO is not crashed and is not already under attack. The input gates \textit{IG\_HW}, \textit{IG\_OS}, and \textit{IG\_SW} allow the failures to happen only when the MANO/5GC is not crashed and the 5G/MANO is under attack. The output gates \textit{OG\_HWF4}, \textit{OG\_OSF3}, and \textit{OG\_SWF2} decrease $M_w$ and increase the token in the respective failure place without moving the token in \textit{Under\_Attack}. The output gates \textit{OG\_Adapts}, \textit{OG\_Man\_Adpt}, \textit{OG\_Att\_rec}, and \textit{OG\_unsucc\_att} delete the token in the incoming places but do not increase $M_w$. \textit{OG\_Man\_Adpt} and \textit{OG\_Att\_rec} reset the failed OS and SW instances, as after a crash. The main purpose of these input and output gates is to avoid the use of the tokens that represent the cluster instances in the parts of the model related to security and connectivity. In case of an attack or connectivity failure, the gates instead generate a new token, which will be deleted in case of an unsuccessful attack or after the recovery.

The 5G/MANO is considered to be down if $M_w<K$ or there are tokens in \textit{HW\_Down}, \textit{OS\_Down},  \textit{SW\_Down}, \textit{MN\_5GCconn\_Fail} (only for MANO), or \textit{Attacked}. 
\section{Evaluation and Discussion} \label{sec:eval}

Using the values in Table~\ref{table:values}, we numerically evaluate the unavailability under the proposed model, also in comparison with the baseline model from~\cite{pathirana2023availability}. Note that the numbers given in Table~\ref{table:values} are mean values, mostly taken from the literature and statistical reports~\cite{tola2019network, nencioni2017including, NguyenmedIoT}. However, the difference in actual values can be high, depending on several factors. For example, the time a system is under attack can last from seconds to months~\cite{rege2017using}. To accommodate the impact of the actual values on the results, we use the sensitivity analysis to check on such parameters. As already mentioned, we implement the model and run the evaluation in M\"{o}bius~\cite{clark2001mobius}.

\begin{table*}[t!]

\caption{System unavailability for different redundancy configurations}
\centering
\resizebox{\textwidth}{!}{
\begin{tabular}{ccccccccccc}
\hline
Row No. & $N_R$ & $N_D$ & $N_C$ & $N_H$ & (K,M) 5GC & (K,M) MANO & Baseline~\cite{pathirana2023availability} & Conn-aware & Sec-aware & New Model \\
\hline
R1 &  1  &  1  &  1  &  1   &  (10,10)  &   (10,10)  & \num{1.256e-1} & \num{1.264e-1} & \num{1.269e-1} & \num{1.279e-1} \\
R2 &  2  &  2  &  2  &  2   &  (9,10)   &   (9,10)   & \num{1.021e-3} & \num{1.115e-3} & \num{1.266e-3} & \num{1.353e-3} \\
R3 &  3  &  3  &  3  &  3   &  (8,10)   &   (8,10)  & \num{1.247e-4} & \num{2.115e-4} & \num{4.503e-4} & \num{4.503e-4} \\
R4 &  4  &  4  &  4  &  4   &  (7,10)   &   (7,10)  & \num{1.136e-4} & \num{2.004e-4} & \num{4.393e-4} & \num{4.393e-4} \\
R5 &  3  &  3  &  3  &  2   &  (9,10)   &   (9,10)   & \num{1.020e-3} & \num{1.115e-3} & \num{1.266e-3} & \num{1.353e-3}   \\
R6 &  2  &  2  &  2  &  3   &  (9,10)   &   (9,10)   & \num{1.020e-3} & \num{1.115e-3} & \num{1.266e-3} & \num{1.353e-3}   \\
R7 &  2  &  2  &  2  &  2   &  (8,10)   &   (8,10)   & \num{1.248e-4} & \num{2.115e-4} & \num{4.504e-4} & \num{4.505e-4} \\
R8 &  3  &  3  &  3  &  2   &  (8,10)   &   (9,10)   & \num{5.727e-4} & \num{6.637e-4} & \num{8.149e-4} & \num{9.020e-4} \\
R9 &  2  &  2  &  2  &  3   &  (9,10)   &   (8,10)   & \num{5.728e-4} & \num{6.631e-4} & \num{9.016e-4} & \num{9.017e-4} \\

\hline
\end{tabular}
}
    \label{table:redundancy}
\end{table*}

Table~\ref{table:redundancy} compares, under different redundancy configurations, the system unavailability by using the model in~\cite{pathirana2023availability} (\emph{Baseline}) and our complete model (\emph{New Model}), as well as each of the two extensions considered separately: the connectivity extension (\emph{Conn-aware}) and the security extension (\emph{Sec-aware}). 

First, we use Table~\ref{table:redundancy} to compare the unavailability values computed using the different models.
As expected, the \emph{Conn-aware} model, which adds the possibility of connectivity failures to the baseline model~\cite{pathirana2023availability}, reduces the system availability. The reduction is minimal for most of the configurations anyway.

The \emph{Sec-aware} model, which adds the possibility of attacks (against both the elements and their connectivity) to the baseline model~\cite{pathirana2023availability}, has a similar behavior than the \emph{Conn-aware} model but a higher reduction of availability across all configurations, closely matching the availability of the \emph{New Model} in most of the cases. The computed unavailability is on the same order of magnitude for all the versions of the model, but the actual unavailability values change from almost the same value (R1, R2, R5, and R6) to higher impacts (R3, R4, and R7), where the unavailability is doubling from the baseline model to the \emph{Conn-aware} model and further doubling from the \emph{Conn-aware} model to the \emph{Sec-aware} model. Naturally, the unavailability is higher in the proposed model since it includes both attacks and connectivity failures as causes of the system outage. The closer results of the \emph{Sec-aware} model to the \emph{New Model} appear natural, as targeted attacks usually are more damaging than unintentional faults, fading their impact. This reflects the crucial role security has in assessing system availability.

Then, we use Table~\ref{table:redundancy} to understand the impact of different redundancies on the unavailability. We see that all the versions of the model have a similar trend. The unavailability is two orders of magnitude higher in R1 (around $10^{-1}$) than in R2, R5, and R6 (around $10^{-3}$) and is around three orders of magnitude higher than in R3, R4, R7, R8, and R9 (around $10^{-4}$).

This behavior highlights that for all the models, a three-element redundancy (R3) is required to reach an unavailability of around $10^{-4}$. A four-element redundancy would not bring a significant further reduction in unavailability. This outcome on the redundancy configuration is important because it indicates the best deployment solution that balances the cost and complexity of additional redundancy versus the incremental gain in availability.

The redundancy configurations R5-R9 show the behavior of the system unavailability when the redundancy is independently incremented for the various sets of elements.
The behavior under different redundancy settings is similar for all models: the redundancy of RAN (R5) or MEHs (R6) has a negligible impact on the system unavailability. Instead, the redundancy of both 5GC and the MANO (R7) has a high impact (around one order of magnitude). 
The redundancy of only one between 5GC and MANO (R8 and R9) also has a noticeable impact on the availability. The importance of 5GC and MANO was already highlighted by the baseline model~\cite{pathirana2023availability}. In conclusion, the best configuration is R7, which has an unavailability that is very close to R3 and R4 but at a lower cost due to the lower number of redundant elements.

\begin{figure}[t!]
    \centering
    \includegraphics[width=\columnwidth]{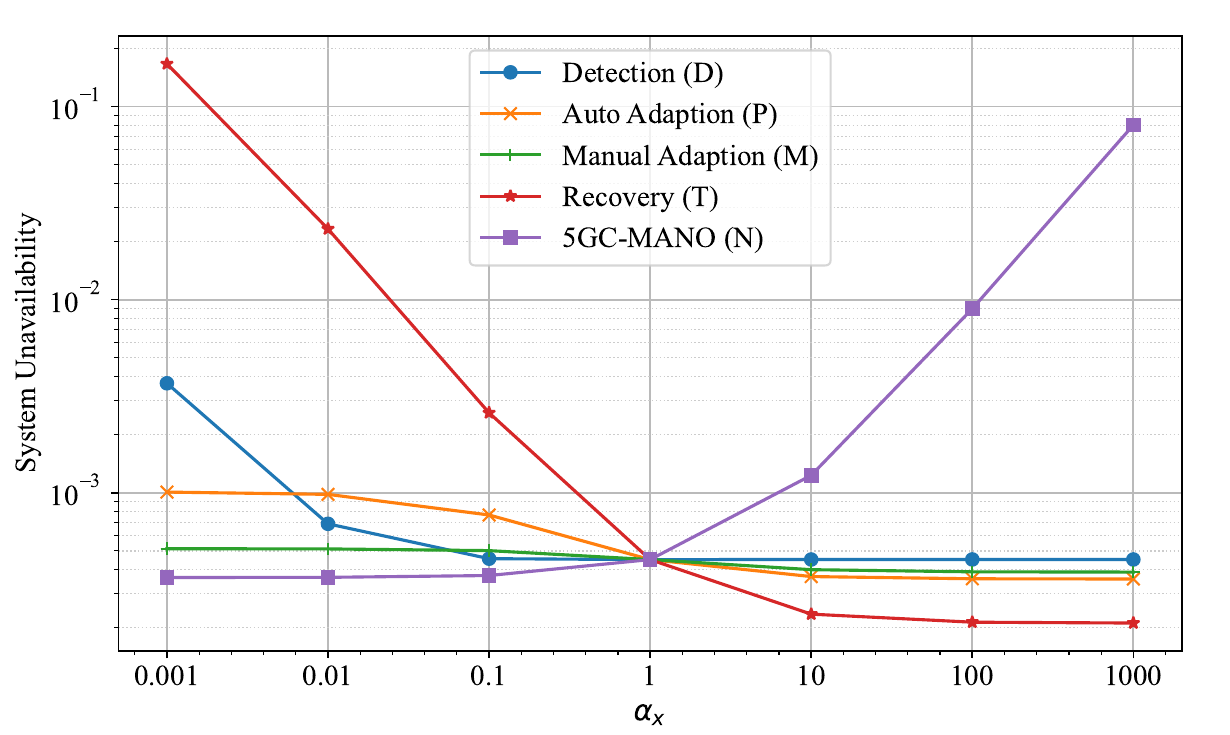}
    \caption{System unavailability by varying the $\alpha$s for detection, automatic and manual adaptions, recovery, and 5GC-MANO} 
    \label{fig:eval:sensec}
\end{figure}

As a result of the above finding, we focus our sensitivity analysis on the security and connectivity extensions of the 5GC and the MANO. Figure~\ref{fig:eval:sensec} shows the system unavailability when the rates of the detection, manual and automatic adaptations, recovery of both the 5GC and the MANO and the failure rate of the 5GC-MANO connection are varied by means of the multiplicative factors $\alpha_X$, $X\in\{D,P,M,T,N\}$. The figure suggests that improving detection and recovery from attacks is essential, as the corresponding times can significantly impact the system availability. Note that while a slow recovery mechanism dramatically increases unavailability, a fast recovery (threshold ca. $\alpha_T = 10$) has a more limited impact. Finally, the figure highlights that the availability of the 5GC-MANO connection, being a single point of failure, highly impacts the overall system availability (threshold ca. $\alpha_N = 0.1$). A bad connection can reduce the availability by a couple of orders of magnitude.

\begin{figure}[b!]
    \centering
    \includegraphics[width=0.95\columnwidth]{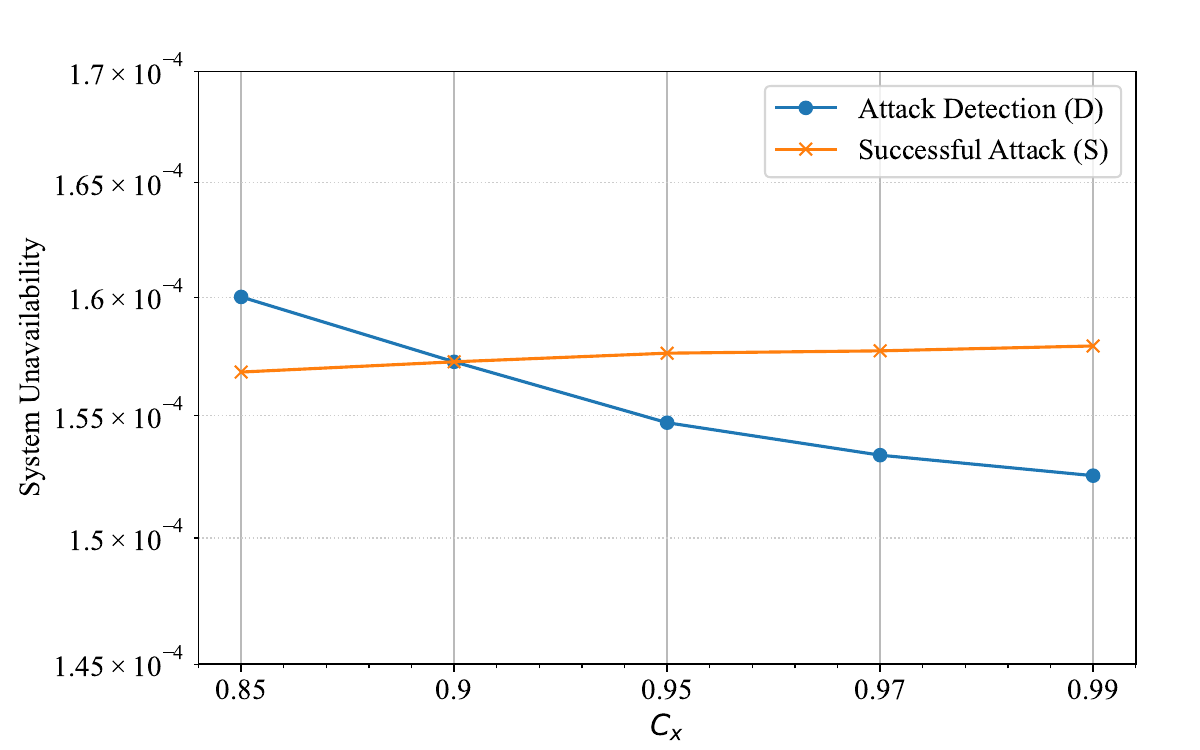}
    \caption{System unavailability varying probability $C$ for attack detection and successful attack of MANO} 
    \label{fig:eval:C_X}
\end{figure}

For $\alpha_T=10$ and $\alpha_N=0.1$, the unavailability reaches $\num{1.572e-4}$. In this configuration, we evaluate the effect of the probabilities to detect and respectively to succeed an attack. Figure~\ref{fig:eval:C_X} shows a low variation in the unavailability when $C_D$ and $C_S$ are varied. 

\begin{figure}[b!]
    \centering
    \includegraphics[width=0.95\columnwidth]{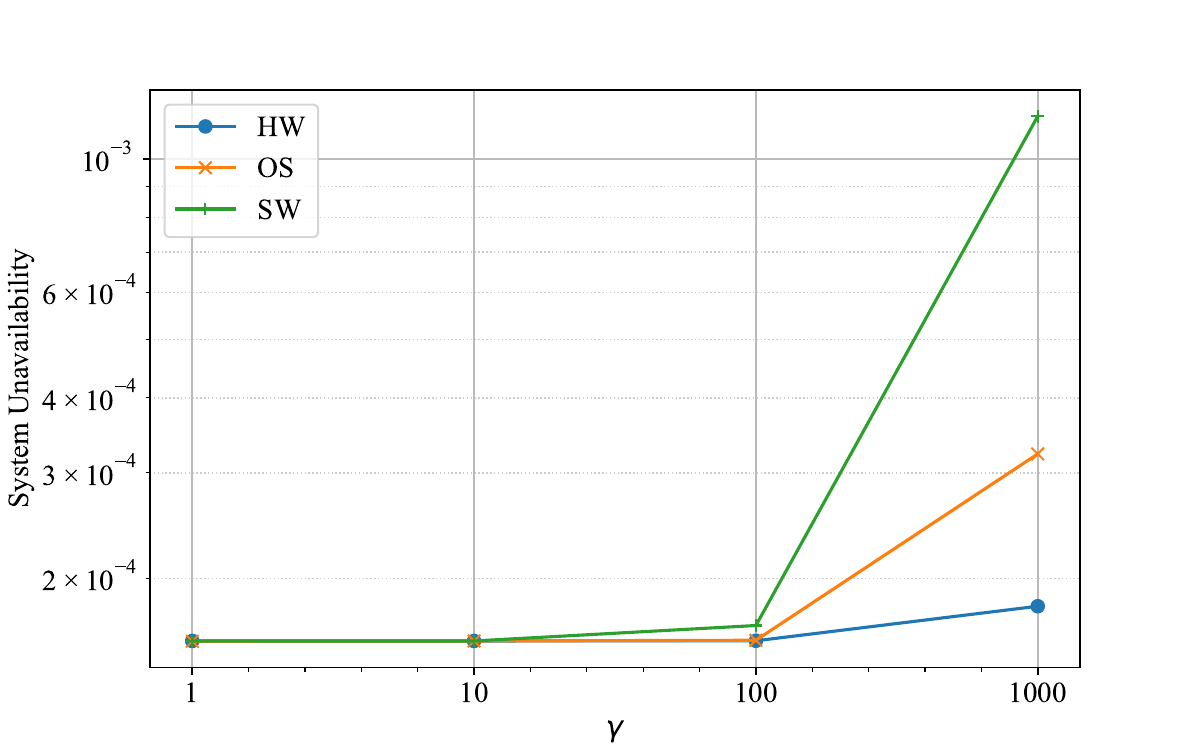}
    \caption{System unavailability varying $\gamma$ for HW, OS and SW  of MANO} 
    \label{fig:eval:sensen}
\end{figure}

Similarly, we evaluate the variation of the amplification factor $\gamma$ for each of the components of the 5GC/MANO: HW, OS, and SW. Figure~\ref{fig:eval:sensen} shows no significant variation from no amplification ($\gamma=1$) to an amplification by two orders of magnitude ($\gamma=100$), except the amplification of the SW failure that is still minimal. Even when three orders of magnitude amplify the failure rates, the variation of the unavailability remains within one order of magnitude for the SW and lower for the HW and the OS.

Our evaluations indicate that although increased overall redundancy enhances the system availability, achieving the 5-nines target requires further optimization. Nevertheless, obtaining the 5-nines redundancy in practical implementation using open-source solutions is usually not feasible or aimed for. 
\section{Conclusion} \label{sec:conclude}

We have proposed connectivity and security extensions for an existing model that evaluates the overall 5G-MEC system availability. Our model maintains the two-level approach of the initial model and defines extended SAN models to accommodate the new dimensions. We found that connectivity failures and security issues do not have a major impact on the 5G-MEC system unavailability being on the same order of magnitude, but they are not negligible because they may lead to an availability reduction of four times. The results confirm the findings in~\cite{pathirana2023availability} regarding the importance of having redundant 5GC and MANO. The impact of the redundancy at the edge (RAN and MEHs) is instead limited. 
The edge already benefits from some redundancy by default (e.g., coverage overlap). In our settings (uncorrelated, independent attacks and faults), this fact impacts both the dependability and security findings (regardless of the fact that the access/edge is more exposed to attacks than the 5GC/MANO). A sensitivity analysis has also highlighted the high impact of a bad connection and of bad attack detection and recovery on the 5G-MEC system availability. The automatic and manual adaptations have instead a more limited impact.

\section*{CRediT author statement}
\textbf{Thilina Pathirana:} Conceptualization, Methodology, Software, Formal analysis, Investigation, Writing - Original Draft, Visualization. \textbf{Gianfranco Nencioni:} Conceptualization, Methodology, Formal analysis, Writing - Review \& Editing, Supervision, Project administration, Funding acquisition. \textbf{Ruxandra F. Olimid:} Conceptualization, Methodology, Writing - Review \& Editing, Supervision.

\bibliographystyle{elsarticle-num} 
\bibliography{}

\end{document}